\documentclass[seceq]{ptptex}
\usepackage{wrapft}
\usepackage{graphicx}

\def\bsub{\begin{subequations}}
\def\esub{\end{subequations}}
\def\beq{\begin{eqnarray}}
\def\eeq{\end{eqnarray}}
\def\bsub{\begin{subequations}}
\def\esub{\end{subequations}}
\def\b{\begin{equation}}
\def\bs{\begin{split}}
\def\es{\end{split}}
\def\e{\end{equation}}

\begin{document}

\title{Spin polarization in high density quark matter\\
under a strong external magnetic field}

\author{Yasuhiko {\sc Tsue}$^{1,2}$, {Jo\~ao da {\sc Provid\^encia}}$^{1}$, {Constan\c{c}a {\sc Provid\^encia}}$^{1}$,\\
{Masatoshi {\sc Yamamura}}$^{1,3}$ and {Henrik {\sc Bohr}}$^{4}$
}

\inst{$^{1}${CFisUC, Departamento de F\'{i}sica, Universidade de Coimbra, 3004-516 Coimbra, 
Portugal}\\
$^{2}${Physics Division, Faculty of Science, Kochi University, Kochi 780-8520, Japan}\\
$^{3}${Department of Pure and Applied Physics, 
Faculty of Engineering Science,\\
Kansai University, Suita 564-8680, Japan}\\
$^{4}${Department of Physics, B.307, Danish Technical University, DK-2800 Lyngby, Denmark}
}

\abst{
In high density quark matter under a strong external magnetic field, 
possible phases are investigated by using the two-flavor Nambu-Jona-Lasinio model with tensor-type 
four-point interaction between quarks, as well as the axial-vector-type four-point interaction.  
In the tensor-type interaction under the strong external magnetic field, it is shown that a quark spin polarized phase is realized 
in all regions of the quark chemical potential under consideration within 
the lowest Landau level approximation. 
In the axial-vector-type interaction, it is also shown that the quark spin polarized phase appears in the wide range of the quark chemical potential. 
In both the interactions, the quark mass in zero and small chemical potential regions increases which indicates that the chiral symmetry breaking is enhanced, namely the magnetic catalysis 
occurs.
}


\maketitle

\section{Introduction}

One of recent interests in the field of the quantum chromodynamics (QCD) may be to understand 
the phase structure under various external environments such as high baryon density, finite temperature, 
finite isospin chemical potential, external magnetic field and so on \cite{FH}. 
In the region with low temperature and large quark chemical potential, it has been remarked that various phases 
may appear such as the color superconducting phase \cite{ARW,IB,CFL}, 
the quarkyonic phase \cite{McL}, the inhomogeneous chiral condensed phase \cite{NT}, 
the quark ferromagnetic phase \cite{Tatsumi}, the color-ferromagnetic phase \cite{Iwazaki}, 
the spin polarized phase due to the axial vector interaction \cite{NMT,TMN} 
or due to the tensor interaction \cite{BJ,IJMP,oursPTP,oursPTEP1,oursPTEP2,oursPTEP3,oursPTEP4} 
and so forth. 
Especially, in these low temperature and large quark chemical potential region 
in the temperature ($T$) and quark chemical potential ($\mu$) plane, it is interesting to investigate 
possible phases in quark matter in the context of the physics of compact star objects such as neutron stars. 
Further, the existence of very strong magnetic field has been found in a certain kind of compact star which is called 
magnetar \cite{magnetar1,magnetar2}. 
On the other hand, in the ultrarelativistic heavy-ion collisions, it has beem remarked 
that a strong magnetic field may be created in the early stage of nucleus-nucleus collisions \cite{KMW}, 
for example, $|eB|\sim m_{\pi}^2$ at the Relativistic Heavy-Ion Collider (RHIC) experiment at Brookhaven, where $e$, $B$ and $m_{\pi}$ represent the elementary electric charge, the magnetic 
flux density or magnetic field and pion mass, respectively, and maybe even stronger at the Large Hadron Collider (LHC) experiment 
at CERN. 
Thus, the investigation of quark matter under an external magnetic field is one of interesting and important problems of QCD 
\cite{Review1,Review2}, especially, it may be very interesting to investigate the basic aspects of quark matter 
under a very strong magnetic field. 

As for the physics of the strong interacting matter under a magnetic field, many investigations are carried out recently \cite{text}. 
For example, the quark mass gap under a strong magnetic field was investigated in the lowest Landau level approximation \cite{Kojo1} 
and in the renormalization-group approach to include the effects of the higher Landau level \cite{Kojo2}, respectively. 
As for the lowest Landau level approximation, the quark gas model under a strong external magnetic field was investigated in order to 
study the massive hybrid compact star object.\cite{sotani} 
By using the $su(3)$ Nambu-Jona-Lasinio model under a strong magnetic field, strange quark matter was also investigated.\cite{Wen}
Recently, much interest is paid to understanding the inhomogeneous chiral condensed phase 
under a magnetic field by using the familiar Nambu-Jona-Lasinio (NJL) model. \cite{Buballa2,Frolov,yoshiike,nishiyama} 
Further, by using the holographic QCD, the spatially-modulated inhomogeneous chiral condensate was also investigated \cite{fukushima}.

Also, in Ref. \citen{fukushima}, the effects of the axial-vector interaction under a strong magnetic field on the spatially-modulated chiral condensed phase 
was investigated by means of the holographic technique. 
On the other hand, the effects of the strong magnetic field on the spin polarized phase has not been investigated widely 
on the extended NJL model with the axial-vector-type four-point interaction between quarks. 
In a preceding work, the effects of the strong magnetic field in the model with the tensor-type interaction have been investigated 
in one-flavor NJL model at finite temperature with zero chemical potential \cite{Ferrer} and the tensor-type interaction has been also used to 
analyze the induced magnetic moment in the context of the magnetic catalysis \cite{Ferrer2}. 
Nevertheless, it is interesting to investigate a possible phase under the strong magnetic field with the tensor-type interaction between quarks 
in two-flavor case at finite quark chemical potential because 
the spin polarized phase
may appear in the region with large quark chemical potential. \cite{oursPTEP4}

In this paper, we investigate possible phases under a very strong external magnetic field 
in the finite quark chemical potential region or in high density quark matter at zero temperature by using the Nambu-Jona-Lasinio (NJL) model \cite{NJL,Buballa,Klevansky,HK} 
with the tensor-type \cite{IJMP} and the axial-vector-type \cite{NMT} four-point interaction between quarks  
as an effective model of QCD. 
As for the tensor-type interaction, we investigate a possible phase under a strong external magnetic field in the case of two flavors and finite quark chemical potential, 
while there is a pioneer work \cite{Ferrer} in the case of one-flavor and zero chemical potential.  
Considering the very strong magnetic field, we focus only on the lowest Landau level to clarify qualitative aspects in quark matter 
under the strong external magnetic field.

This paper is organized as follows: 
In the next section, a simple model as an extension of the original NJL model is introduced which reveals qualitative aspects 
of quark matter. 
In Sect. 3, the contribution of the tensor-type interaction is investigated, which is a main purpose of this paper. 
As a result, it is shown that the spin polarized phase under the strong external magnetic field exists. 
In Sect. 4, the contribution of the axial-vector-type interaction is investigated and 
whether a spin polarized phase appears or not is discussed. 
Also, we point out the implication of the axial-vector condensate to the spatially-modulated inhomogeneous chiral condensate in the Appendix. 
The last section is devoted to the summary and concluding remarks.

\setcounter{equation}{0} 
\section{Mean field approximation for the Nambu-Jona-Lasinio model with tensor-type and/or vector-axial vector-type four-point interactions between quarks
}

Let us start from the two-flavor Nambu-Jona-Lasinio model with tensor-type \cite{BJ,IJMP,Nova} and vector-axial vector-type \cite{NMT,TMN} 
four-point interactions between quarks 
under the external magnetic field. 
The Lagrangian density can be expressed as 
\beq\label{2-1}
{\cal L}&=&
{\bar \psi}(i\gamma^\mu D_{\mu}-m_0)\psi+G_s[({\bar \psi}\psi)^2+({\bar \psi}i\gamma_5{\vec \tau}\psi)^2]\nonumber\\
& &-\frac{G_p}{2}[({\bar \psi}\gamma^\mu{\vec \tau}\psi)^2+({\bar \psi}i\gamma_5\gamma^\mu{\vec \tau}\psi)^2]
-\frac{G_T}{4}[({\bar \psi}\gamma^{\mu}\gamma^{\nu}{\vec \tau}\psi)^2+({\bar \psi}i\gamma_5\gamma^{\mu}\gamma^{\nu}\psi)^2]\ ,
\eeq
where $m_0$ represents a current quark mass and $D_\mu$ represents the covariant derivative introduced as 
\beq\label{2-2}
D_\mu=\partial_\mu+iQA_\mu\ ,\qquad 
A_\mu=\left(0,\ \frac{By}{2},\ -\frac{Bx}{2},\ 0\right) = (0,\ -{\mib A})\  .
\eeq
Here, $Q=2e/3$ for up quark and $-e/3$ for down quark are the electric charges where $e$ is the elementary charge.
There is an external magnetic field $B$ along $z$-axis.  

Hereafter, we treat the model within the mean field approximation. 
In order to consider the spin polarization under the mean field approximation, 
the axial-vector condensate $\langle {\bar \psi}\gamma_5\gamma^3\tau_3\psi\rangle$ and 
the tensor condensate $\langle {\bar \psi}\gamma^1\gamma^2\tau_3\psi\rangle$ are taken into account. 
Then, the Lagrangian density is reduced into 
\beq\label{2-3}
{\cal L}_{MF}&=&{\bar \psi}(i\gamma^\mu D_\mu -M_q)\psi+U_A{\bar \psi}\gamma_5\gamma^3\tau_3\psi
+F_3{\bar \psi}i\gamma^1\gamma^2\tau_3\psi-\frac{M^2}{4G_s}-\frac{U_A^2}{2G_p}-\frac{F_3^2}{2G_T}
\nonumber\\
&=&{\bar \psi}(i\gamma^\mu D_\mu -M_q)\psi-U_A{\psi^{\dagger}}\Sigma_3\tau_3\psi
-F_3{\bar \psi}\Sigma_3\tau_3\psi-\frac{M^2}{4G_s}-\frac{U^2}{2G_p}-\frac{F^2}{2G_T} \ , 
\eeq
where 
\beq\label{2-4}
& &\Sigma_3=-i\gamma^1\gamma^2=-\gamma^0\gamma_5\gamma^3=
\left(
\begin{array}{cc}
\sigma_3 & 0 \\
0 & \sigma_3
\end{array}
\right)
\nonumber\\
& &M_q=m_0+M\ , \qquad M=-2G_s\langle{\bar \psi}\psi\rangle\ , \nonumber\\
& &U_A=G_p\langle {\bar \psi}\gamma_5\gamma^3\tau_3\psi\rangle=-G_p\langle\psi^{\dagger}\Sigma_3\tau_3\psi\rangle\equiv U\tau_f\ , 
\nonumber\\
& &
F_3=iG_T\langle{\bar \psi}\gamma^1\gamma^2\tau_3\psi\rangle=-G_T\langle{\bar \psi}\Sigma_3\tau_3\psi\rangle=F\tau_f\ . 
\eeq
Here, $\tau_f=1$ for up quark and $-1$ for down quark, which denote the eigenvalues of $\tau_3$. 
Also, $\sigma_3$ is the third component of the Pauli spin matrices.

Introducing the quark chemical potential $\mu$ in order to consider finite density quark matter, 
the Hamiltonian density can be obtained from the Lagrangian density within the mean field approximation as 
\beq\label{2-5}
{\cal H}_{MF}-\mu{\cal N}
&=&{\bar \psi}\left(-i{\mib \gamma}\cdot ({\mib \nabla}-iQ{\mib A})+M_q-\mu\gamma^0+U_A\gamma^0\Sigma_3\tau_3
+F_3\Sigma_3\tau_3\right)\psi
\nonumber\\
& &
+\frac{M^2}{4G_s}+\frac{U^2}{2G_p}+\frac{F^2}{2G_T}\ ,
\eeq
where ${\cal N}$ represents the quark number density, $\psi^{\dagger}\psi$.

Let us treat the two cases separately in the following sections. 
One is a possibility of the existence of only the tensor condensation without axial-vector condensate. 
Another is the axial-vector condensation without the tensor condensate.

\setcounter{equation}{0}

\section{The case of the tensor condensation}

In this section, possible phases are investigated in the case of the tensor-type four-point interaction between quarks. 
It was found that the spin polarized phase may be realized above a certain critical quark chemical potential 
without an external magnetic field \cite{oursPTEP1,oursPTEP4}. 
Thus, it is interesting to investigate whether the spin polarized phase is realized or not under a strong external magnetic field.

\subsection{Thermodynamic potential}

In the case of the tensor condensation without axial-vector condensate, the Hamiltonian density 
can be written as 
\beq\label{4-1}
{\cal H}_{MF,T}-\mu{\cal N}
&=&{\bar \psi}\left(-i{\mib \gamma}\cdot ({\mib \nabla}-iQ{\mib A})+M_q-\mu\gamma^0
+F_3\Sigma_3\tau_3\right)\psi
+\frac{M^2}{4G_s}+\frac{F^2}{2G}\nonumber\\
&=&\psi^{\dagger}(h_T-\mu)\psi +\frac{M^2}{4G_s}+\frac{F^2}{2G_T}\ .
\eeq
Here, $h_T$ can be obviously defined. 
In order to obtain the eigenvalues of $h_T$, namely the energy eigenvalues of a single quark, it is necessary to diagonalize $h_T$. 
The detail derivation has already been carried out in Refs. \citen{oursPTEP3} and \citen{oursPTEP4}. 
As a result, the eigenvalues of $h_T$ which are written as $E_{T,p\nu\eta}^{f}$ where $p$, $\nu$ and $f$ represent 
the $z$-component of momentum, the quantum number of Landau level and flavor, respectively, are obtained: 
\beq\label{4-2}
E_{T,p\nu\eta}^{f}
&=&
\left\{
\begin{array}{l}
{\displaystyle 
E_{T,p\nu\sigma}^u=\sqrt{\left(F+\sigma\sqrt{M_q^2+2Q_u B\nu}\right)^2+p_z^2}}\ , \\
{\displaystyle 
\qquad\qquad\qquad\qquad\qquad\qquad\qquad\ \ 
\left\{
\begin{array}{ll}
\nu=0,1,2,\cdots\ & {\rm for}\ \ \sigma=1 \\
\nu=1,2,\cdots \ & {\rm for}\ \ \sigma=-1
\end{array}
\right.}\\
{\displaystyle
E_{T,p\nu\sigma}^d=\sqrt{\left(-F+\sigma\sqrt{M_q^2-2Q_d B\nu}\right)^2+p_z^2}}\ ,\\
{\displaystyle 
\qquad\qquad\qquad\qquad\qquad\qquad\qquad\ \  
\left\{
\begin{array}{ll}
\nu=1,2,\cdots & {\rm for}\ \ \sigma=1 \\
\nu=0,1,2,\cdots  & {\rm for}\ \ \sigma=-1
\end{array}
\right.}
\end{array}\right.
\nonumber\\
&=&
\sqrt{\left(F+\eta\sqrt{M_q^2+2|Q_f|B\nu}\right)^2+p_z^2}\ , 
\quad
\left\{
\begin{array}{ll}
\nu=0, 1,2,\cdots & {\rm for}\ \  \eta=1 \\
\nu=1,2,\cdots & {\rm for}\ \ \eta=-1
\end{array}
\right.
\qquad
\eeq 
where $Q_f=2e/3$ for $f=u$ and $-e/3$ for $f=d$, respectively. 

The thermodynamic potential has been also given in Ref.\citen{oursPTEP3}. Including the contribution of vacuum, 
the thermodynamic potential $\Phi_T$ can be expressed  as 
\beq\label{4-3}
\Phi_T&=&
\sum_{f,\alpha}\int^{\Lambda}\frac{dp_z}{2\pi}\frac{|Q_f|B}{2\pi}\left(E_{T,p\ \nu=0\ \eta=1}^f-\mu\right)\theta(\mu-E_{T,p\ \nu=0\ \eta=1}^f)
\nonumber\\
& &
+
\sum_{\eta,f,\alpha}\int^{\Lambda}\frac{dp_z}{2\pi}\sum_{\nu=1}^{E_T<\mu}\frac{|Q_f|B}{2\pi}\left(E_{T,p\nu\eta}^f-\mu\right)
\theta(\mu-E_{T,p\nu\eta}^f)
\nonumber\\
& &-\sum_{f,\alpha}\int^{\Lambda}\frac{dp_z}{2\pi}\frac{|Q_f|B}{2\pi}E_{T,p\ \nu=0\ \eta=1}^f
-
\sum_{\eta,f,\alpha}\int^{\Lambda}\frac{dp_z}{2\pi}\sum_{\nu=1}^{E_T<\Lambda}\frac{|Q_f|B}{2\pi}E_{T,p\nu\eta}^f
\nonumber\\
& &+\frac{M^2}{4G_s}+\frac{F^2}{2G_T}\ ,
\eeq 
where $\alpha$ represents the quark color and $\Lambda$ represents the three-momentum cutoff. 
Here, $\theta(x)$ represents the Heaviside step function.
The first and second lines represent the positive-energy contribution of quarks and the third line represents 
the vacuum contribution.

\subsection{A possible phase under a strong external magnetic field}

In this subsection, possible phases under a 
strong external magnetic field at zero temperature and finite chemical potential are investigated in the case of the tensor-type interaction between quarks. 
Under a strong magnetic field, only the lowest Landau level plays an essential role 
because the higher Landau levels with $\nu\neq 0$ 
have no contribution to the thermodynamic potential due to $E_{T,p\nu\eta}^f > \mu$ or 
the three-momentum cutoff $\Lambda$.
Namely, 
the effective model under consideration is valid for the region in which 
the energy or the magnitude of three-momentum is less than the three-momentum 
cutoff $\Lambda$. 
Thus, if $\sqrt{2|Q_f|B} \geq \Lambda$, only the lowest Landau level with $\nu=0$ contributes 
to the thermodynamic potential. 
Let us consider the system in the above-mentioned situation. 
Hereafter, we set $m_0=0$ in the chiral limit for simplicity and then $M_q=M$.

Since we omit the higher Landau level with $\nu\neq 0$, the thermodynamic potential 
(\ref{4-3}) is reduced into 
\beq\label{4-4}
\Phi_T&\approx&
\sum_{f,\alpha}\int^{\Lambda}\frac{dp_z}{2\pi}\frac{|Q_f|B}{2\pi}\left(E_{T,p\ \nu=0\ \eta=1}^f-\mu\right)\theta(\mu-E_{T,p\ \nu=0\ \eta=1}^f)
\nonumber\\
& &
-\sum_{f,\alpha}\int^{\Lambda}\frac{dp_z}{2\pi}\frac{|Q_f|B}{2\pi}E_{T,p\ \nu=0\ \eta=1}^f
+\frac{M^2}{4G_s}+\frac{F^2}{2G_T}
\nonumber\\
&=&-\frac{3eB}{4\pi^2}\left[
\mu\sqrt{\mu^2-(F+M)^2}
-(F+M)^2\ln\frac{\mu+\sqrt{\mu^2-(F+M)^2}}{F+M}\right]
\nonumber\\
& &\qquad\ \ 
\times\theta(\mu-(M+F))
\nonumber\\
& &-\frac{3eB}{4\pi^2}\left[\Lambda\sqrt{\Lambda^2+(F+M)^2}
+(F+M)^2\ln\frac{\Lambda+\sqrt{\Lambda^2+(F+M)^2}}{F+M}\right]
\nonumber\\
& &
+\frac{M^2}{4G_s}+\frac{F^2}{2G_T}\ . 
\eeq 
The gap equations for $M$ and $F$ are derived from the relations 
${\partial \Phi_T}/{\partial M}=0$ and ${\partial \Phi_T}/{\partial F}=0$. 
Here, we can safely neglect the derivative term of the Heaviside step function with respect to $M$ and $F$, which leads to a Dirac delta function, because 
the delta function is different from zero only for $\mu=M+F$. 
This divergence is an artifact in calculation. 
Thus, the gap equations are given as 
\beq\label{4-5}
& &\frac{\partial \Phi_T}{\partial M}=\frac{M}{2G_s}+I_0(M,F)+I_\mu(M,F)=0\ ,
\nonumber\\
& &\frac{\partial \Phi_T}{\partial F}=\frac{F}{G_T}+I_0(M,F)+I_\mu(M,F)=0 \ ,
\eeq
where 
\beq\label{4-5add}
& &I_0(M,F)=-\frac{3eB}{2\pi^2}(F+M)\ln\frac{\Lambda+\sqrt{\Lambda^2+(F+M)^2}}{F+M}\ , 
\nonumber\\
& &I_\mu(M,F)=\frac{3eB}{2\pi^2}(F+M)\ln\frac{\mu+\sqrt{\mu^2-(F+M)^2}}{F+M}\theta(\mu-(F+M))\ . 
\eeq
Then, we obtain the following celebrated relation between the chiral condensate $M$ and the spin polarized condensate $F$ again: 
\beq\label{4-6}
M=\frac{2G_s}{G_T}F\ ,
\eeq
which was first derived by Ferrer et al. \cite{Ferrer} in the case of zero chemical potential in one-flavor NJL model.   
In the case of no external magnetic field, the chiral condensate, $M$, and the spin polarized condensate, $F$, are independent from 
each other \cite{oursPTEP4}. 
Namely, (i) in the region with small quark chemical potential, the chiral condensate is not equal to zero, but the 
spin polarized condensate does not appear, namely, the chiral broken phase, $M\neq 0$ and $F=0$, is realized. 
(ii) In the intermediate range of the quark chemical potential, the chiral symmetric phase with $M=F=0$ is realized. 
(iii) In the region with large quark chemical potential, the spin polarized phase, $M=0$ and $F\neq 0$, is realized. 
On the other hand, under a strong external magnetic field under investigation, 
the chiral condensate and the spin polarized condensate are related to 
each other through the relation (\ref{4-6})  even for a finite quark chemical potential. 
It is an interesting feature originated from the external magnetic field.

Eliminating $M$ by using the relation (\ref{4-6}), then, the gap equation for $F$ in (\ref{4-5}) is recast into 
\beq
& &\mu\leq F+M \nonumber\\
& &\qquad
F\left[\frac{1}{G_T}-\frac{3eB}{2\pi^2}\left(1+\frac{2G_s}{G_T}\right)
\ln\frac{\Lambda+\sqrt{\Lambda^2+\left(1+\frac{2G_s}{G_T}\right)^2F^2}}{\left(1+\frac{2G_s}{G_T}\right) F}\right]=0\ , 
\label{4-7}\\
& &\mu>F+M \nonumber\\
& &\qquad
F\left[
\frac{1}{G_T}+
\frac{3eB}{2\pi^2}\left(1+\frac{2G_s}{G_T}\right)\ln\frac{\mu+\sqrt{\mu^2-F^2\left(1+\frac{2G_s}{G_T}\right)^2}}{\Lambda+\sqrt{\Lambda^2+F^2
\left(1+\frac{2G_s}{G_T}\right)^2}}\right]=0\ . 
\label{4-8}
\eeq 
It is understood that the trivial solution, $F=0$, always exists.

For $\mu \leq F+M$, the solution of gap equation with $F\neq 0$ is obtained analytically as 
\beq
& &F=\frac{G_T\Lambda}{G_T+2G_S}\cdot\frac{1}{\sinh\frac{2\pi^2}{3eB(G_T+2G_s)}}\ , 
\label{4-7-2}
\\
& &M=\frac{2G_s\Lambda}{G_T+2G_S}\cdot\frac{1}{\sinh\frac{2\pi^2}{3eB(G_T+2G_s)}}\ .
\nonumber
\eeq

\begin{table}[b]
\caption{ Parameter set for the cases of the tensor-type interaction and the axial-vector interaction }
\begin{center}
{\begin{tabular}{ccc} \hline
$eB=0.597\ {\rm GeV}^2$ & $2G_s=2G_p=G_T=11.0\ {\rm GeV}^{-2}$ & $\Lambda=0.631\ {\rm GeV}$ \\
\hline
\end{tabular}}
\end{center}
\end{table}

\begin{figure}[t]
\begin{center}
\includegraphics[height=5.5cm]{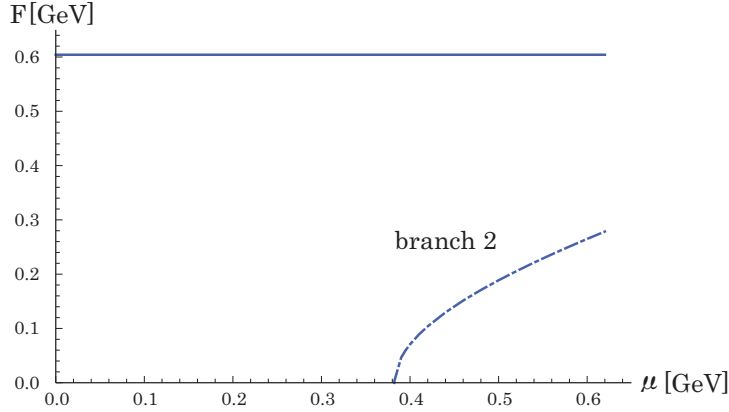}
\caption{Spin polarized condensate $F$ is depicted as a function of the quark 
chemical potential $\mu$. 
}
\label{fig:fig3}
\end{center}
\end{figure}

Let us give numerical results simply. 
We take the parameters $2G_s=G_T=11.0$ GeV$^{-2}$, $eB=0.597 $ GeV$^2$ ($\approx m_{\rho}^2$) and $\Lambda=0.631$ GeV, 
where $m_{\rho}$ represents the rho meson mass.  
As for $G_s$ and $\Lambda$, these are taken as standard values \cite{HK}, which give the dynamical quark mass in vacuum as 
$M=M_q+m_0=0.335$ GeV 
with the current quark mass $m_0\equiv m_u=m_d=0.005$ GeV or $M_0\equiv M=0.322$ GeV with $m_0=0$ in the chiral limit. 
In these parameters,  $\sqrt{2|Q_f|B}\geq \Lambda$ is satisfied. 
Thus, only the lowest Landau level contributes to the thermodynamic potential. 
The parameters under consideration are summarized in Table 1. 
With these parameters, the non-trivial solution for $\mu \leq F+M$ in (\ref{4-7-2}) always exists. 
The results are as follows:
\beq\label{4-7-3}
F=0.604\ {\rm GeV}\ , \qquad\quad
M=0.604\ {\rm GeV}\ . \qquad\quad
(\mu \leq \Lambda < F+M)
\eeq
In addition to the above solution,  
it is shown that there exists a non-trivial solution of the gap equation (\ref{4-8}) for $\mu > F+M$ above the quark chemical potential 
$\mu=\mu_{T,{\rm cr}}=0.382$ GeV. 
In Fig.\ref{fig:fig3}, the spin polarized condensates $F$ 
are depicted by solid line for $\mu \leq F+M$ and by dash-dotted curve for $\mu > F+M$ denoted by ``branch 2" 
as a function of the quark chemical potential $\mu$.

%
%

%
\begin{figure}[t]
\begin{center}
\includegraphics[height=5.5cm]{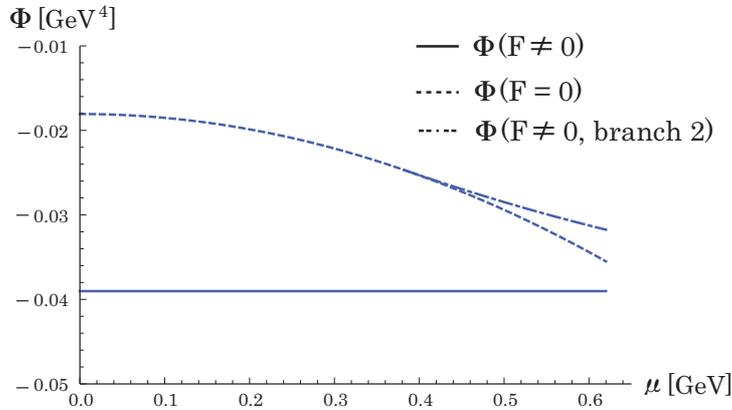}
\caption{The thermodynamic potentials with non-zero $F\ (\neq 0)$ (solid line), 
the second branch with non-zero $F$ (dash-dotted curve) above $\mu_{T,{\rm cr}}$ and $F=0$ (dotted curve) 
are depicted as a function of the quark chemical potential $\mu$.
}
\label{fig:fig4}
\end{center}
\end{figure}
%

Of course, there always exists the trivial solution $F=0$. 
It is necessary to compare the thermodynamic potential of $F=0$ with that of $F\neq 0$ in order to determine 
which phase is realized. 
In Fig.\ref{fig:fig4}, the thermodynamic potentials with non-zero $F(\neq 0)$ (solid line) in the case of $\mu\leq F+M$, 
non-zero $F$ (dash-dotted curve) in the second branch with $\mu\geq \mu_{T,{\rm cr}} ( >F+M) $ and 
that of the trivial 
solution with $F=0$ (dotted curve) are shown. 
From this figure, it is seen that the thermodynamic potential with constant spin polarized condensate 
is always lower than those of the other cases of the second branch $F\neq 0$ and of $F=0$.  
Thus, the spin polarized condensate appears in all ranges of the quark chemical potential under consideration. 
Therefore, within the tensor-type four-point interaction between quarks, 
the spin polarized phase is realized. 
Of course, in this spin polarized phase, the dynamical quark mass has the value $M=0.604$ GeV in (\ref{4-7-3}), which is larger than 
the value $M_0=0.322$ GeV that is the dynamical quark mass in vacuum without the external magnetic field.
Therefore, the chiral symmetry breaking is enhanced due to the external magnetic field, namely, the magnetic catalysis actually occurs.

\setcounter{equation}{0}
\section{The case of the axial-vector condensation}

In this section, the case of the axial-vector condensation without tensor condensate is considered.

\subsection{Thermodynamic potential} 
The Hamiltonian density 
can be written as 
\beq\label{3-1}
{\cal H}_{MF,A}-\mu{\cal N}
&=&{\bar \psi}\left(-i{\mib \gamma}\cdot ({\mib \nabla}-iQ{\mib A})+M_q-\mu\gamma^0
+U_A\gamma^0\Sigma_3\tau_3\right)\psi
+\frac{M^2}{4G_s}+\frac{U^2}{2G_p}\nonumber\\
&=&\psi^{\dagger}(h_A-\mu)\psi +\frac{M^2}{4G_s}+\frac{U^2}{2G_p}\ .
\eeq
Here, $h_A$ can also be defined obviously in (\ref{3-1}). 
In order to obtain the eigenvalues of $h_A$, namely the energy eigenvalues of a single quark, it is necessary to diagonalize $h_A$. 
Similarly to the derivation of the eigenvalues of $h_T$, the eigenvalues of $h_A$ can be obtained easily.  
As a result, the eigenvalues of $h_A$ which are written as $E_{A,p\nu\eta}^{f}$ 
are obtained: 
\beq\label{3-2}
E_{A,p\nu\eta}^{f}&=&
\left\{
\begin{array}{l}
{\displaystyle 
E_{A,p\nu\sigma}^u=\sqrt{2Q_u B\nu+\left(\sqrt{p_z^2+M_q^2}+\sigma U\right)^2}}\ ,\\
{\displaystyle 
\qquad\qquad\qquad\qquad\qquad\qquad\qquad\ \  
\left\{
\begin{array}{ll}
\nu=0,1,2,\cdots\ & {\rm for}\ \ \sigma=1 \\
\nu=1,2,\cdots \ & {\rm for}\ \ \sigma=-1
\end{array}
\right.}\\
{\displaystyle
E_{A,p\nu\sigma}^d=\sqrt{-2Q_d B\nu+\left(\sqrt{p_z^2+M_q^2}-\sigma U\right)^2}}\ ,\\
{\displaystyle
\qquad\qquad\qquad\qquad\qquad\qquad\qquad\ \  
\left\{
\begin{array}{ll}
\nu=1,2,\cdots & {\rm for}\ \ \sigma=1 \\
\nu=0,1,2,\cdots  & {\rm for}\ \ \sigma=-1
\end{array}
\right.}
\end{array}\right. \qquad
\nonumber\\
&=&
\sqrt{2|Q_f|B\nu+\left(\eta U+\sqrt{p_z^2+M_q^2}\right)^2}\ .
\quad
\left\{
\begin{array}{ll}
\nu=0, 1,2,\cdots & {\rm for}\ \  \eta=1 \\
\nu=1,2,\cdots & {\rm for}\ \ \eta=-1
\end{array}
\right.
\eeq 

The thermodynamic potential has been also calculated as is similar to the case of the tensor condensation without 
axial-vector condensate in the preceeding section. 
The results is as follows: 
\beq\label{3-3}
\Phi_A&=&
\sum_{f,\alpha}\int^{\Lambda}\frac{dp_z}{2\pi}\frac{|Q_f|B}{2\pi}\left(E_{A,p\ \nu=0\ \eta=1}^f-\mu\right)
\theta(\mu-E_{A,p\ \nu=0\ \eta=1}^f)
\nonumber\\
& &
+
\sum_{\eta,f,\alpha}\int^{\Lambda}\frac{dp_z}{2\pi}\sum_{\nu=1}^{E_A<\mu}\frac{|Q_f|B}{2\pi}\left(E_{A,p\nu\eta}^f-\mu\right)
\theta(\mu-E_{A,p\nu\eta}^f)
\nonumber\\
& &-\sum_{f,\alpha}\int^{\Lambda}\frac{dp_z}{2\pi}\frac{|Q_f|B}{2\pi}E_{A,p\ \nu=0\ \eta=1}^f
-
\sum_{\eta,f,\alpha}\int^{\Lambda}\frac{dp_z}{2\pi}\sum_{\nu=1}^{E_A<\Lambda}\frac{|Q_f|B}{2\pi}E_{A,p\nu\eta}^f
\nonumber\\
& &
+\frac{M^2}{4G_s}+\frac{U^2}{2G_p}\ . 
\eeq 
The first and the second lines represent the positive-energy contribution of quarks and the third line represents 
the vacuum contribution. 
It should be noted that the single quark energy does not depend on the flavor in the lowest Landau level with $\nu=0$.

\subsection{A possible phase under the strong external magnetic field}

In this subsection, possible phases under a 
strong external magnetic field at zero temperature and finite chemical potential are investigated in the case of the axial-vector-type interaction between quarks 
within the lowest Landau level approximation. 
Namely, if $\sqrt{2|Q_f|B} \geq \Lambda$, only the lowest Landau level with $\nu=0$ has the contribution 
to the thermodynamic potential.

Since we omit the higher Landau level with $\nu\neq 0$, the thermodynamic potential 
(\ref{3-3}) is reduced into 
\beq\label{3-13}
\Phi_A&\approx&
\sum_{f,\alpha}\int^{\Lambda}\frac{dp_z}{2\pi}\frac{|Q_f|B}{2\pi}\left(E_{A,p\ \nu=0\ \eta=1}^f-\mu\right)\theta(\mu-E_{A,p\ \nu=0\ \eta=1}^f)
\nonumber\\
& &-\sum_{f,\alpha}\int^{\Lambda}\frac{dp_z}{2\pi}\frac{|Q_f|B}{2\pi}E_{A,p\ \nu=0\ \eta=1}^f
+\frac{M^2}{4G_s}+\frac{U^2}{2G_p}
\nonumber\\
&=&\frac{3eB}{4\pi^2}\biggl[
-(\mu-U)\sqrt{(\mu-U)^2-M^2}
+M^2\ln\frac{\mu-U+\sqrt{(\mu-U)^2-M^2}}{M}\biggl]
\nonumber\\
& &\qquad\qquad
\times\theta(\mu-(U+M))
\nonumber\\
& &
-\frac{3eB}{4\pi^2}\biggl[\Lambda\sqrt{\Lambda^2+M^2}+M^2\ln\frac{\Lambda+\sqrt{\Lambda^2+M^2}}{M}+2U\Lambda
\biggl]
+\frac{M^2}{4G_s}+\frac{U^2}{2G_p}\ . \ \ 
\eeq 
The gap equations are separately derived in the case $\mu \leq U+M$ and $\mu > U+M$, respectively, as 
\beq
& &\mu \leq U+M \nonumber\\
& &\qquad
\frac{\partial \Phi_A}{\partial M}=M\left[-\frac{3eB}{2\pi^2}\ln\frac{\Lambda+\sqrt{\Lambda^2+M^2}}{M}+\frac{1}{2G_s}\right]=0\ ,
\nonumber\\
& &\qquad
\frac{\partial \Phi_A}{\partial U}=-\frac{3eB}{2\pi^2}\Lambda+\frac{U}{G_p}=0 \ . 
\label{3-15}\\
& &\mu>U+M \nonumber\\
& &\qquad
\frac{\partial \Phi_A}{\partial M}=M\left[\frac{3eB}{2\pi^2}\ln\frac{\mu-U+\sqrt{(\mu-U)^2-M^2}}{\Lambda+\sqrt{\Lambda^2+M^2}}+\frac{1}{2G_s}\right]=0\ ,
\nonumber\\
& &\qquad
\frac{\partial \Phi_A}{\partial U}=\frac{3eB}{2\pi^2}\left[\sqrt{(\mu-U)^2-M^2}-\Lambda\right]+\frac{U}{G_p}=0 \ . 
\label{3-14}
\eeq
For the case $\mu \leq U+M$, the gap equations (\ref{3-15}) are solved easily and the solutions are obtained as 
\beq\label{3-17}
& &
M=
\left\{
\begin{array}{l}
0 \\
{\displaystyle 
\frac{\Lambda}{\sinh\left(\frac{\pi^2}{3eBG_s}\right)} \ (\equiv M_A)}\\
\end{array}\right.
\\
& &
U=\frac{3eBG_p}{2\pi^2}\Lambda\ , \nonumber
\eeq
and the thermodynamic potential under the above solutions is obtained in both the cases, $M=0$ and $M=M_A$, as 
\beq\label{3-17-2}
& &\Phi_A=
\left\{
\begin{array}{ll}
{\displaystyle \Phi_A^<(M=0)=-\frac{3eB}{4\pi^2}\Lambda^2\left(1+\frac{3eBG_p}{2\pi^2}\right)} 
\\
{\displaystyle 
\Phi_A^<(M=M_A) = \frac{M_A^2}{4G_s}}\\
{\displaystyle \qquad \qquad -\frac{3eB}{4\pi^2}\left(\Lambda\sqrt{\Lambda^2+M_A^2}+M_A^2\ln\frac{\Lambda+\sqrt{\Lambda^2+M_A^2}}{M_A}+\frac{3eBG_p}{2\pi^2}\Lambda^2\right)}\ . 
\end{array}\right. \qquad
\eeq
For the case $\mu > U+M$, the solutions for (\ref{3-14}) are written as 
\beq\label{3-16}
& &
M=
\left\{
\begin{array}{l}
0 \\
{\displaystyle 
\frac{2\sqrt{C}}{1+C^2}\sqrt{C\left((\mu-U)^2-\Lambda^2\right)+\Lambda(1-C^2)(\mu-U)} \ (\equiv M_C)}\\
\end{array}\right.\\
& &
U=\frac{3eBG_p}{2\pi^2-3eBG_p}(\Lambda-\mu)\ , \nonumber
\eeq
where ${\displaystyle C=\exp\left(-\frac{\pi^2}{3eBG_s}\right)}$ and the thermodynamic potential is obtained as 
\beq\label{3-16-2}
& &\Phi_A=
\left\{
\begin{array}{ll}
{\displaystyle \Phi_A^>(M=0)=-\frac{3eB}{4\pi^2}\left[(\mu-U)^2+\Lambda^2+2U\Lambda\right]+\frac{U^2}{2G_p}} 
\\
{\displaystyle 
\Phi_A^>(M=M_C) = -\frac{3eB}{4\pi^2}\biggl[(\mu-U)\sqrt{(\mu-U)^2-M_C^2}+\Lambda\sqrt{\Lambda^2+M_C^2}}\\
\qquad\qquad\qquad\qquad\qquad\ \ 
{\displaystyle +2U\Lambda
-M_C^2\ln\frac{\mu-U+\sqrt{(\mu-U)^2-M_C^2}}{\Lambda+\sqrt{\Lambda^2+M_C^2}}\biggl]}\\
\qquad\qquad\qquad\qquad\qquad
{\displaystyle 
+\frac{M_C^2}{4G_s}+\frac{U^2}{2G_p}}\ .
\end{array}\right. \qquad
\eeq

Let us give numerical results simply. 
We take parameters $eB=0.597$ GeV$^2$, $G_s=G_p=5.5$ GeV$^{-2}$ and $\Lambda=0.631$ GeV which are given in Table 1. 
With these parameters,  $\sqrt{2|Q_f|B}\geq \Lambda$ is satisfied, so, only the lowest Landau level with $\nu=0$ contributes to the thermodynamic 
potential. 
In these parameters, 
for $\mu \leq U+M$, the quark mass $M$, spin polarized condensate $U$ and the thermodynamic potential $\Phi_A^<$ do not depend on $\mu$. 
Namely, 
\beq\label{4-1}
& &M=0\ , \qquad {\rm or}\qquad 
M=M_A=0.536\ {\rm GeV} \ , 
\nonumber\\
& &U=0.315\ {\rm GeV}\ , \nonumber\\
& &\Phi_A=
\left\{
\begin{array}{l}
\Phi_A^<(M=0)=-0.0271\ {\rm GeV}^4 \\
\Phi_A^<(M=M_A)=-0.0327\ {\rm GeV}^4
\end{array}\right.
\eeq
On the other hand, for $\mu>U+M$, the non-trivial solution with $M\neq 0$ appears above $\mu=\mu_{A,{\rm cr}}=0.431$ GeV.
The dynamical quark masses $M$ are depicted as a function of the quark chemical potential $\mu$ in Fig.3. 
Of course, for any value of chemical potential, there exists the trivial solution $M=0$.

%
\begin{figure}[t]
\begin{center}
\includegraphics[height=5.5cm]{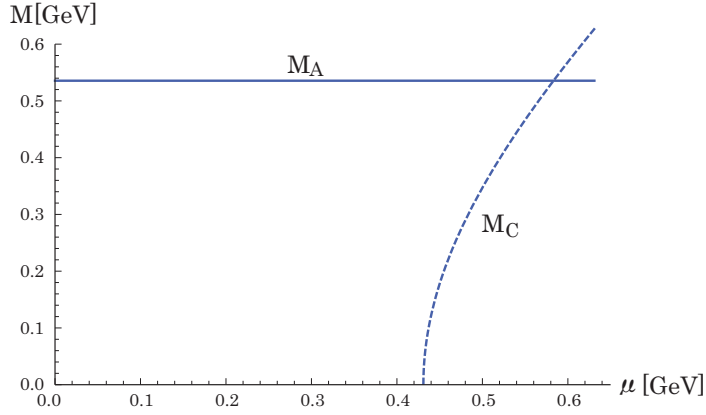}
\caption{The dynamical quark mass $M$ are depicted as a function of the quark chemical potential $\mu$. 
For $\mu \leq U+M$, the non-trivial solution $M_A$ always appears. 
For $\mu>U+M$, the non-trivial solution $M_C$ appears above a certain critical chemical potential 
$\mu_{A,{\rm cr}}=0.431$ GeV. 
}
\label{fig:fig1}
\end{center}
\end{figure}
%

%
\begin{figure}[t]
\begin{center}
\includegraphics[height=5.5cm]{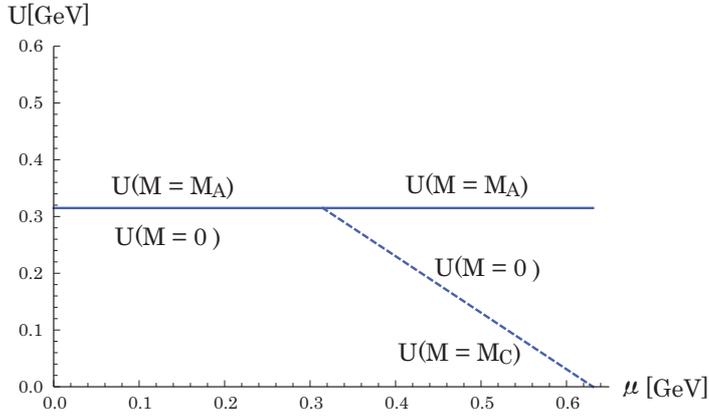}
\caption{The spin polarized (axial-vector) condensate $U$ is shown as a function of quark 
chemical potential $\mu$. 
}
\label{fig:fig1}
\end{center}
\end{figure}
%

%
\begin{figure}[t]
\begin{center}
\includegraphics[height=5.5cm]{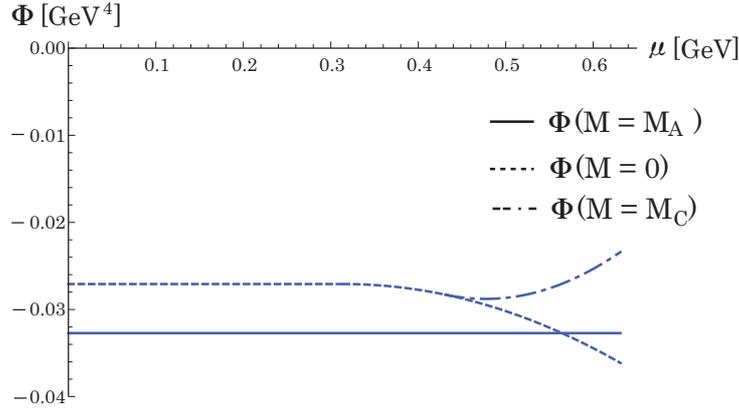}
\caption{The thermodynamic potentials with $M=M_A$ (solid line), $M=M_C$ (dash-dotted curve) and $M=0$ (dotted curve) 
are depicted as a function of the quark chemical potential $\mu$.
}
\label{fig:fig2}
\end{center}
\end{figure}
%

Also, in Fig.4, the spin polarized condensate $U$ is shown as a function of $\mu$. In all regions for the chemical potential, 
the non-trivial solution, $U=0.315$ GeV, with $M=M_A$ exists. 
With $M=0$, $U$ is constant below $\mu=\mu_1=0.315$ GeV. Above $\mu_1$, $U$ decreases linearly with respect to $\mu$. 
Above $\mu=\mu_{A,{\rm cr}}$, another branch appears with $M=M_C$, but the value is the same as $U$ with $M=0$.

It is necessary to compare the thermodynamic potentials in order to determine 
which phase is realized with $M=0$, $M_A$ or $M_C$. 
In fig.\ref{fig:fig2}, the thermodynamic potentials with $M=M_A$ (solid line), $M=0$ (dotted curve) and $M=M_C$ (dash-dotted curve) above $\mu_{A,{\rm cr}}$ 
are shown as functions of 
the quark chemical potential $\mu$. 
In all regions of $\mu$, the spin polarized condensate $U$ dose not zero due to the strong external magnetic field. 
Also, in almost regions for $\mu$, the realized phase has the non-zero dynamical quark mass. 
Thus, the realized phase may be the spin polarized phase. 
Of course, $M=M_A=0.536$ GeV $ > M_0=0.322$ GeV, where $M_0$ is the vacuum value of the dynamical quark mass 
in this parameter set used here. 
Thus, the magnetic catalysis occurs similarly to the case of the tensor-type interaction. 
From this figure, it is indicated within this model and parameter set that, 
at large quark chemical potential about 0.564 GeV, the phase change may occur form the phase with $M=M_A$ 
to $M=0$.

\setcounter{equation}{0}
\section{Summary and concluding remarks}

It was investigated whether there exists a phase with the spin polarization in high density quark matter 
due to the tensor-type or the axial-vector-type interaction between quarks under a strong external magnetic field or not. 
In this paper, under the assumption of a strong magnetic field, only the contribution of the lowest Landau level was investigated. 

In the case of the tensor-type four-point interaction between quarks in the NJL model, 
the tensor condensate, namely spin polarized condensate 
$F=-G_T\langle {\bar \psi}\Sigma_3\tau_3\psi\rangle\tau_f$, appears in the whole region of the quark chemical potential $\mu$ under consideration. 
However, there is always the trivial solution $F=0$ for the gap equation for any value of the quark chemical potential. 
The realized phase in high density quark matter is determined by comparing the thermodynamic potential for $F\neq 0$ with
that for $F=0$. 
It is found that the realized phase is the phase with $F\neq 0$, namely, there is spin polarization even when there is a strong external magnetic
field. 
In the whole region of the chemical potential, the dynamical quark mass is larger than the vacuum value. 
Therefore, the magnetic catalysis actually occurs under the existence of the tensor-type interaction.

In the case of the axial-vector-type four-point interaction between quarks in the NJL model, 
there exists the axial-vector condensate, namely spin polarization $U=-G_p\langle \psi^{\dagger}\Sigma_3\tau_3\psi\rangle\tau_f$, 
for any value of the quark chemical potential under the strong external magnetic field. 
The axial-vector condensate represents the expectation value of the spin matrix $\Sigma_3$ with respect to the 
quark number density $\psi^{\dagger}\psi$. 
Under a strong magnetic field, this expectation value always has a finite value as is expected. 
Although the spin polarization $U$ has non-zero value in all region with respect to the quark chemical potential $\mu$, 
the quark mass or the chiral condensate, $M$, has a finite value or zero in various region of $\mu$. 
However, up to $\mu=0.564$ GeV for the parameter set used in this paper, 
the spin polarized phase with $M=M_A \neq 0$ is realized. 
The axial-vector condensate is related to the spatially-modulated inhomogeneous chiral condensate, which 
is described in Appendix briefly. 
Thus, the knowledge of the inhomogeneous chiral condensed phase may be useful to investigate the 
spin polarized condensate originated from the axial-vector-type interaction considered in this paper.

In this paper, we have investigated the possible phases in high density quark matter 
under a strong external magnetic field in both the cases of the existence of the tensor-type 
four-point interaction and that of the axial-vector-type four-point interaction between quarks. 
It is interesting to study the system under the external magnetic field in which there exist 
the effects of the higher Landau level with $\nu\neq 0$, while in this paper only 
the contribution of the lowest Landau level has been taken into account by considering the 
very strong external magnetic field. 
The spin polarization may lead to the spontaneous magnetization of quark matter \cite{oursPTEP3}.
It is also interesting to study the implication to the origin of the strong magnetic field appearing in the compact stars such as neutron star or magnetar.\cite{harding}
In this paper, the quark chemical potentials for $u$- and $d$-quarks are taken as the same values $\mu$, so 
the system is not charge neutral and is not in $\beta$-equilibrium because we have investigated symmetric quark matter. 
When we consider the high density quark matter in the inner core of the compact stars, 
the above conditions for the quark chemical potentials are demanded. 
These are important future problems in this study.

\section*{Acknowledgment}

One of the authors (Y.T.) would like to express their sincere thanks to\break
Professor J. da Provid\^encia and Professor C. Provid\^encia, two of co-authors of this paper, 
for their warm hospitality during their visit to Coimbra in spring of 2016. 
One of the authors (Y.T.) 
is partially supported by the Grants-in-Aid of the Scientific Research 
(No.26400277) from the Ministry of Education, Culture, Sports, Science and 
Technology in Japan.

\vspace{-0cm}

\appendix

\section{Implication to the inhomogeneous chiral condensate}

The axial-vector-type spin polarized condensate is closely related to the inhomogeneous chiral condensate. 
It is shown in this appendix that the axial-vector condensed phase is identical with the spatially-modulated inhomogeneous chiral condensed phase  
by means of the unitary transformation \cite{Frolov,BJ}.

The Lagrangian density of the original NJL model is written as 
\beq\label{3-4}
{\cal L}_{NJL}={\bar \psi}i\gamma^\mu \partial_{\mu}\psi 
+G_s\left[({\bar \psi}\psi)^2+({\bar \psi}i\gamma_5{\vec \tau}\psi)^2\right]\ . 
\eeq
Under the mean field approximation, in general, the above Lagrangian density is recast into 
\beq\label{3-5}
{\cal L}_{NJL}^{MF}&=&
{\bar \psi}\left[i\gamma^{\mu}\partial_{\mu}+2G_s\left(
\sigma+{\vec \pi}i\gamma_5{\vec \tau}\right)\right]\psi
-\sigma^2-{\vec \pi}^2\ , 
\eeq
where, under the existence of the inhomogeneous chiral condensation, the condensations $\sigma$ and ${\vec \pi}$ 
can be expressed as 
\beq\label{3-6}
& &\sigma\equiv \langle{\bar \psi}\psi\rangle={\bar \sigma}\cos {\mib q}\cdot{\mib r}\ , 
\nonumber\\
& &{\vec \pi}\equiv \langle{\bar \psi}i\gamma_5{\vec \tau}\psi\rangle=
\left\{
\begin{array}{ll}
0 & {\rm for}\ \  \pi_1=\pi_2\\
{\bar \sigma}\sin{\mib q}\cdot{\mib r} \qquad & {\rm for}\ \ \pi_3 \ .  
\end{array}\right.
\eeq
Here, ${\bar \sigma}$ represents the radius of the chiral circle and ${\mib q}$ is a certain 
vector characterizing the inhomogeneous chiral condensate. 
The condensations depend on the space coordinate ${\mib r}$, which leads to the spatially-modulated inhomogeneous chiral condensate. 
The Hamiltonian is easily derived:
\beq\label{3-7}
& &
H_{NJL}^{MF}=\int d^3{\vec r}\ {\cal H}_{NJL}^{MF}
=\int d^3{\vec r}\ \psi^{\dagger}h_{NJL}^{MF}\psi +V\left(\sigma^2+{\vec \pi}^2\right)\ , 
\nonumber\\
& &
h_{NJL}^{MF}=-i{\mib \alpha}\cdot{\mib \nabla}-2G_s\beta\left(\sigma+i\gamma_5{\vec \pi}\cdot{\vec \tau}\right)\ .
\eeq
Here, $V$ represents the volume under consideration and the gamma matrices are taken as 
\beq\label{3-8}
{\mib \alpha}=\gamma^0{\mib \gamma}
=\left(
\begin{array}{cc}
0 & {\mib \sigma} \\
{\mib \sigma} & 0
\end{array}
\right)\ , \quad
\beta=\gamma^0
=\left(
\begin{array}{cc}
1 & 0 \\
0 &-1
\end{array}
\right)\ , \quad
\gamma_5
=\left(
\begin{array}{cc}
0 & 1 \\
1 & 0
\end{array}
\right)\ , \qquad
\eeq
where ${\mib \sigma}$ represent the $2\times 2$ Pauli matrices. 

Under the inhomogeneous chiral condensate in Eq.(\ref{3-6}), 
the Hamiltonian matrix $h_{NJL}^{MF}$ can be re-expressed as 
\beq\label{3-9}
h_{NJL}^{MF}&=&
-i{\mib \alpha}\cdot{\mib \nabla}-2G_s\beta{\bar \sigma}\left(\cos{\mib q}\cdot{\mib r}+i\gamma_5\tau_3\sin{\mib q}\cdot{\mib r}
\right)\nonumber\\
&=&
-i{\mib \alpha}\cdot{\mib \nabla}-2G_s\beta{\bar \sigma}e^{i\gamma_5\tau_3{\mib q}\cdot{\mib r}}
\nonumber\\
& =&-i{\mib \alpha}\cdot{\mib \nabla}-2G_s{\bar \sigma}
e^{-\frac{i}{2}\gamma_5\tau_3{\mib q}\cdot{\mib r}}\beta e^{\frac{i}{2}\gamma_5\tau_3{\mib q}\cdot{\mib r}}\ , 
\eeq
where we used the anticommutation relation $\{\ \beta\ , \ \gamma_5\ \}=0$. 
Therefore, performing a unitary transformation,
we get a unitary-equivalent Hamiltonian as 
\beq\label{3-10}
h'&=&e^{\frac{i}{2}\gamma_5\tau_3{\mib q}\cdot{\mib r}} h_{NJL}^{MF} e^{-\frac{i}{2}\gamma_5\tau_3{\mib q}\cdot{\mib r}}
\nonumber\\
&=&-i{\mib \alpha}\cdot{\mib \nabla}-2G_s{\bar \sigma}\beta -\frac{1}{2}\tau_3{\mib \Sigma}\cdot{\mib q}\ , 
\nonumber\\
{\mib \Sigma}&=&\gamma_5{\mib \alpha}
=\left(
\begin{array}{cc}
{\mib \sigma} & 0 \\
0 & {\mib \sigma}
\end{array}
\right)\ .
\eeq
This procedure corresponds to moving from ``rest frame" to the ``rotating frame" in isospin space, the idea of which 
was introduced to treat the disoriented chiral condensate as a quasi-stable state in relativistic heavy ion collisions 
\cite{TVM1,TVM2}. 
Finally, we obtain 
\beq
& &h\psi=E\psi\ , \qquad {\rm then}\ , \qquad
h'\psi'=E\psi' \ , \nonumber\\
& &h'=-i{\mib \alpha}\cdot{\mib \nabla}+\beta M+U_A\tau_3\Sigma_3\ , 
\label{3-11}\\
& &\qquad M=-2G_s{\bar \sigma}\ , \qquad U_A=-\frac{1}{2}q\qquad {\mib q}=(0,\ 0,\ q)\ , 
\nonumber\\
& &\psi'=e^{\frac{i}{2}\gamma_5\tau_3{\mib q}\cdot{\mib r}}\psi\ .  
\nonumber
\eeq
Here, $h'$ is identical with $h_A$ in Eq.(\ref{3-1}) with ${\mib A}={\mib 0}$.
Thus, the eigenvalue $E$ is easily obtained as 
\beq\label{3-12}
E&=&\sqrt{p_x^2+p_y^2+\left(\sqrt{p_z^2+M^2}+\eta U\right)^2} \nonumber\\
&=&\sqrt{p_x^2+p_y^2+\left(\sqrt{p_z^2+M^2}\pm \frac{q}{2}\right)^2} \ ,
\eeq
where $U\tau_f =U_A$. 
Thus, the axial-vector condensate is identical with the inhomogeneous condensate by means of the 
unitary transformation.

\end{document}